\documentclass[a4paper]{llncs}

\usepackage{epsfig}

\newtheorem{thm}{Theorem}
\newtheorem{lem}{Lemma}

\newtheorem{fig}{Figure}

\def\leurre{\noindent\leftskip0pt\small\baselineskip 10pt}

\def\encadre#1#2{%
\setbox100=\hbox{\kern#1{#2}\kern#1}
\dimen100=\ht100 \advance \dimen100 by #1
\dimen101=\dp100 \advance \dimen101 by #1
\setbox100=\hbox{\vrule height \dimen100 depth \dimen101\box100\vrule}
\setbox100=\vbox{\hrule\box100\hrule}
\advance \dimen100 by .4pt \ht100=\dimen100
\advance \dimen101 by .4pt \dp100=\dimen101
\box100
\relax
}

\def\ligne#1{\hbox to \hsize{#1}}
\def\PlacerEn#1 #2 #3 {\rlap{\kern#1\raise#2\hbox{#3}}}

\title{Is the injectivity of the global function of a cellular automaton in the hyperbolic
plane undecidable?}
\author{Maurice Margenstern\inst{1}}
\institute{%
Laboratoire d'Informatique Th\'eorique et Appliqu\'ee, EA 3097,\\
Universit\'e de Metz, I.U.T. de Metz,\\
D\'epartement d'Informatique,\\
\^Ile du Saulcy,\\
57045 Metz Cedex, France,\\
\email{margens@univ-metz.fr}
}
\begin{document}
\maketitle

\vskip 10pt
\begin{abstract}
In this paper, we look at the following question. We consider cellular automata in
the hyperbolic plane, see \cite{mmJUCSii,mmkmTCS,mmarXivc,mmbook1} and we consider
the global function defined on all possible configurations. Is the injectivity
of this function undecidable? The problem was answered positively in the case
of the Euclidean plane by Jarkko Kari, in 1994, see \cite{jkari94}. In the present paper,
we give a partial answer: when the configurations are restricted to a certain condition,
the problem is undecidable.
\end{abstract}
\vskip 10pt

\def\cqfd{\hbox{\kern 2pt\vrule height 6pt depth 2pt width 8pt\kern 1pt}}
\def\Hii{$I\!\!H^2$}
\def\Hiii{$I\!\!H^3$}
\def\Hiv{$I\!\!H^4$}
\def\norm{\hbox{$\vert\vert$}}
\section{Introduction}

   The global function of a cellular automaton~$A$ is defined in the set of 
all configurations. This is a very different point of view than implementing an
algorithm to solve a given problem. In this latter case, the initial configuration
is usually finite.
  
   In the case of the Euclidean plane, the definition of the set of configurations
is very easy: it is $Q^{Z\!\!Z^2}$, where $Q$~is the set of states of the automaton.

   In the hyperbolic plane, see~\cite{mmbook1}, following what we did in~\cite{mmarXivc}, 
we have the following situation: we consider that the grid is the pentagrid or the ternary
heptagrid, see~\cite{mmbook1}. We fix a tile, which will be called the {\bf central cell} 
and, around it, we dispatch $\alpha$~sectors, $\alpha\in\{5,7\}$: $\alpha=5$ in the case 
of the pentagrid, $\alpha=7$ in the case of the ternary heptagrid. We assume that the
sectors and the central cell cover the plane and the sectors do not overlap, neither 
the central cell, nor other sectors: call them the {\bf basic sectors}. Denote 
by ${\cal F}_\alpha$ the set constituted by the central cell and $\alpha$~Fibonacci 
trees, each one spanning a basic sector. Then, a configuration of a cellular automaton~$A$
in the hyperbolic plane can be represented as an element of~$Q^{{\cal F}_\alpha}$, where
$Q$ is the set of states of~$A$. If $f_A$~denotes the {\bf local} transition function
of~$A$, its {\bf global} transition function~$G_A$ is defined by: $G_A(c)(x)=f(c(x))$.

The injectivity problem for a cellular automaton consists in asking whether there
is an algorithm which, applied to a description of~$A$ would indicate whether 
$G_A$~is injective or not.

   In this paper, we give a partial negative answer to this question: we also assume 
that~$A$ satisfies a constraint which we shall describe in the fourth section.
In order to describe the constraint, we need some preliminary material which we give
in the second section.
 
\section{Preliminary constructions}

   The constructions to which we now turn are described for the ternary heptagrid.
However, this can also be performed in the pentagrid, at a price of a different
setting which, here, we have not the room to present. In the first sub-section, 
we indicate the basis of the construction which is described by the second sub-section. 
 
\subsection{The interwoven triangles}

   The interwoven triangles implement a structure which can be simpler described in the
Euclidean plane. It is illustrated by figure~\ref{interwoven}, below.

\vskip 10pt
\setbox110=\hbox{\epsfig{file=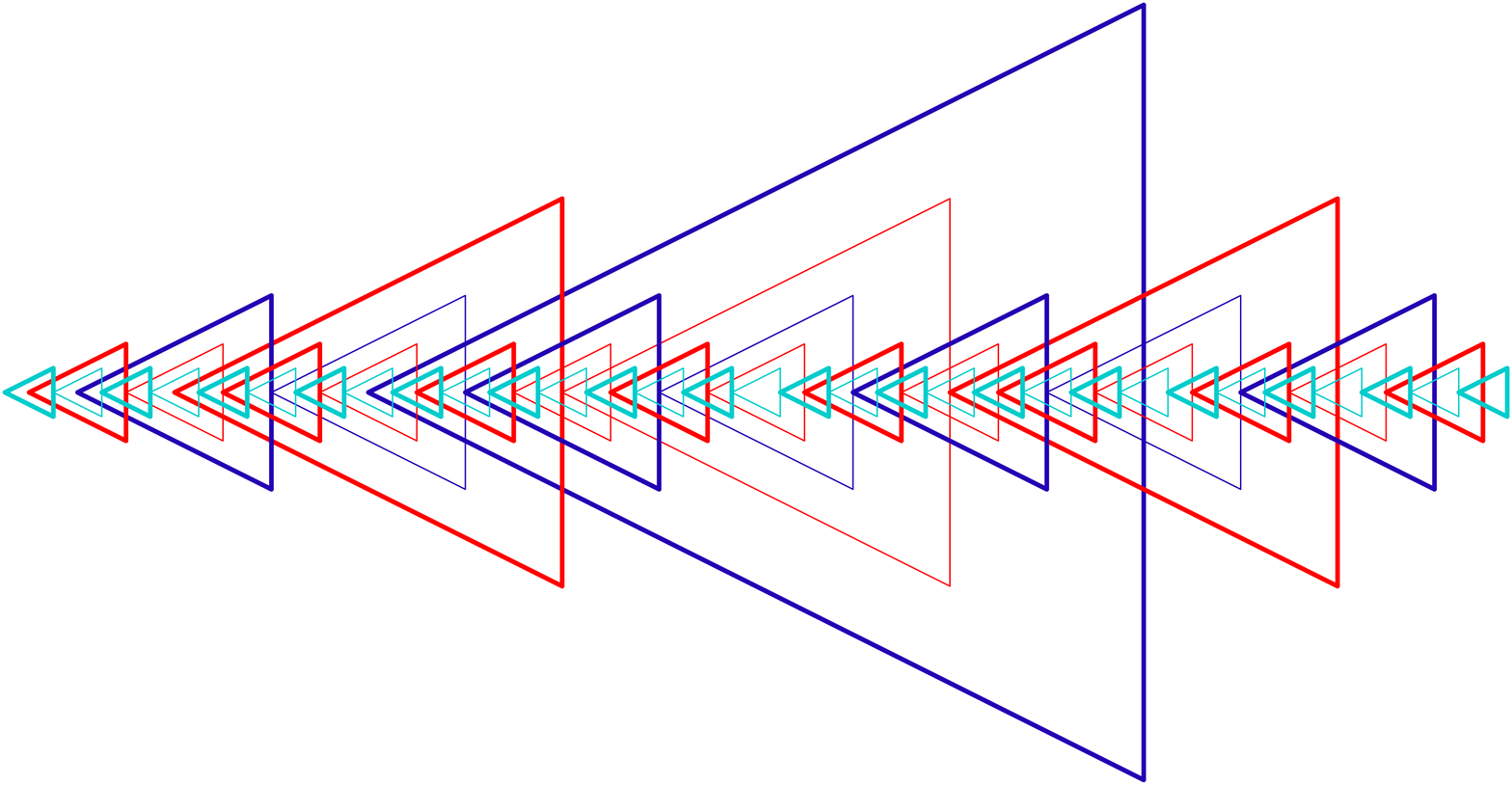,width=300pt}}
\vtop{
\ligne{\hfill
\PlacerEn {-330pt} {0pt} \box110
}
\vskip-15pt
\begin{fig}\label{interwoven}
\leurre
Interwoven triangles in the Euclidean plane.
\end{fig}
}
\vskip 10pt
   We fix a first infinite set of triangles, which we call the generation~0 and which
by definition, have the colour blue-0. They are isosceles triangles with all their
symmetry axes on the same line, called the {\bf axis}. The triangles are all equal 
and the vertex of the one is the mid-point of the basis of the other. Now, each second 
triangle is dotted with thick sides, while the others are equipped with thin ones. From 
now on, call {\bf triangles} of the generation~0 those which have thick sides and 
call {\bf phantoms} of the generation~0, those which have thin sides. As shown in 
figure~\ref{interwoven}, triangles and phantoms alternate: the vertex of a triangle, 
of a phantom, is the mid-point of the basis of a phantom, of a triangle respectively. 
For properties shared by both triangles and phantoms, we speak of {\bf trilaterals}. We 
call {\bf mid-point of the legs} the mid-points of the sides of a trilateral which are 
not its basis. By construction, there is a {\bf green line} between the mid-points of 
the legs of each phantom of the generation~0.

\vskip 10pt
\setbox110=\hbox{\epsfig{file=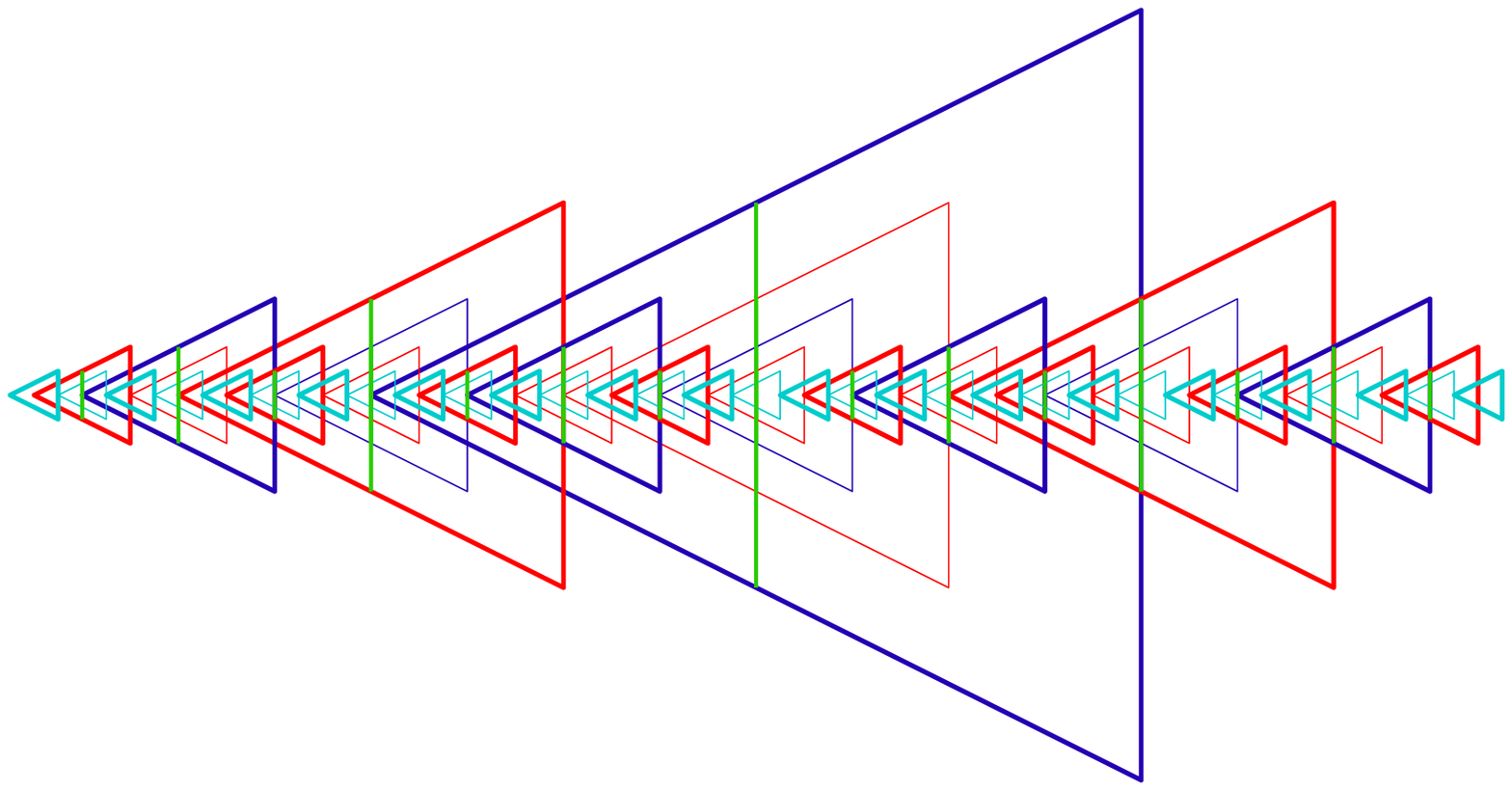,width=400pt}}
\vtop{
\ligne{\hfill
\PlacerEn {-390pt} {0pt} \box110
}
\vskip-45pt
\begin{fig}\label{interwovengreen}
\leurre
Construction of the interwoven triangles in the Euclidean plane: the green signal.
\end{fig}
}
\vskip 10pt

   Now, assume that we have constructed the trilaterals of the generation~$n$. When $n$~is
positive, the colour of the trilaterals of the generation is blue or red and these colours
are said to be {\bf opposite} to each other. We consider that the red colour is also 
opposite to the blue-0 one. When the construction of the trilaterals of the generation~$n$ 
is completed, there is a green line which goes from the mid-point of a leg to the other 
in each trilateral when $n>0$, in each phantom when $n=0$. At this point, we choose 
a triangle~$T$, at random. Then, on the mid-point of the segment which joins the mid-points 
of the legs of~$T$, this segment is the green line of~$T$ when $n>0$, we put the vertex~$V$ 
of an isosceles triangle~$S$ of an opposite colour with respect to that of the 
generation~$n$ and whose legs are parallel to those of~$T$. To the right-hand side
of~$T$, there is a phantom~$P$ of the generation~$n$, whose vertex is the mid-point of the
basis of~$T$. The green line of~$P$ goes on across the mid-points of the legs of~$P$ and
they meet the legs of~$S$ at their mid-point: the legs of~$S$ stop the green line at these
points. Now, from the mid-point of the basis of~$S$, we construct a phantom~$Q$ which
is the image of~$S$ under a shift along the axis whose amplitude is the height of~$S$:
of course, the sides of~$Q$ are thin while those of~$S$ are thick. Then, we reproduce 
the pattern constituted by the union of~$S$ and~$Q$ by shifts along the
axis of amplitude the distance between the vertex of~$S$ and the mid-point of the basis
of~$Q$, the vertex of a triangle, of a phantom being the mid-point of the basis.
The trilaterals~$S$ and~$Q$ and their images under the just indicated shifts constitute
the trilaterals of the generation~$n$+1. This construction is recursively repeated.
The union of all the generations constitute the {\bf interwoven triangles}, also
see figure~\ref{interwovengreen}.
See also~\cite{mmBEATCS,mmarXiva} for a more precise description of this construction
and of its properties. Here are the main properties of the interwoven triangles:

\begin{lem}\label{interwovenproperties} {\rm (Margenstern, see \cite{mmBEATCS,mmarXiva})}
$-$
The trilaterals of the interwoven triangles have the following properties:
\vskip 0pt
{\parindent 0pt
$(i)$ The triangles of the same colour never intersect: their areas are disjoint or one 
of them contains the other. 
\vskip 0pt
$(ii)$ The legs of a trilateral never intersect the leg of another
trilateral. But it may intersect the basis of a few trilaterals.
\par}
\vskip 0pt
More precisely, the rules of these intersections are:
\vskip 0pt
{\parindent 0pt
$(iii)$ The intersection of a leg of a trilateral with the basis of another one only 
occurs between the vertex and the mid-point of the leg.
\vskip 0pt
$(iv)$ A leg of a triangle intersects the basis of a triangle of the previous generation.
\vskip 0pt
$(v)$ A leg of a phantom intersects the basis of exactly one trilateral for all the 
previous generations. The trilateral of the previous generation which is met by the leg
is a triangle and all the other bases which are met by the leg belong to a phantom.
\vskip 0pt
$(vi)$ Phantoms can be grouped into {\bf towers} which are finite sets of phantoms
which have the same green line. The phantoms of a tower have their areas 
successively included one in the other. All generations up to the last one of the
tower are present in the tower.
\par}
\end{lem}
  
   Now, we consider that the axis has marks which are defined by the vertices of the 
trilaterals of the generation~0, as well as the mid-points between the vertex of a 
trilateral~$T$ and the mid-point of the basis~$T$. We say that each mark defines a 
{\bf row} which, by definition is a line which meets trilaterals. Consider a red 
triangle~$R$ and let $\rho$ be a row which cuts the legs of~$R$: we exclude the
rows which contain the vertex or the basis of~$R$. We say that $\rho$~is {\bf free}
in~$R$ if and only if does not meet a red triangle inside~$R$: if $\rho$ meets the
vertex or the basis of a red triangle inside~$R$ it is also not free. Similarly,
we define the free rows of the blue and blue-0 triangles. In \cite{mmBEATCS,mmarXiva},
we prove the following property:

\begin{thm}\label{freerows} {\rm (Margenstern, see \cite{mmBEATCS,mmarXiva})} $-$
In the generation~$2n$$+$$1$, with $n\in I\!\!N$, each red triangle contains 
$2^{n+1}$$+$$1$ free rows exactly. Each blue or blue-0 triangle contains one free row 
exactly: it accompanies the green line in the blue triangles; it crosses the mid-point 
between the vertex and the mid-point of the basis in the blue-0 triangles. 
\end{thm}

\subsection{Implementing the interwoven triangles in the hyperbolic plane}

   As indicated in~\cite{mmBEATCS,mmarXive}, we implement this Euclidean construction
in the hyperbolic plane, namely in a tiling based on the ternary heptagrid, also see 
\cite{mmbook1} for this latter tiling. The construction is based on a particular
tiling of the ternary heptagrid called the {\bf mantilla}. Here, we very sketchily 
indicate the construction of the mantilla. We define it by rules applied to two kinds
of tiles, the {\bf centres} and the {\bf petals}. By construction, each petal must abut
three centers and a centre is surrounded by petals only. A centre together with its
seven petals constitute a {\bf flower}. We mark two kinds of flowers by introducing
one or two red {\bf vertices} at a vertex shared by two petals of the flower, but not any 
centre. Of course, a red vertex is shared by three adjacent flowers. A flower with one 
red vertex only is called an {\bf 8}-flower. There are two kinds of flowers with
two red vertices: the difference depends on the smallest number of petals which
are around the centre, between these two red vertices and which do not share it.
In the $F$-flowers, this number is~0 and in the $G$-flower, this number is~1. We
can also go from one red-vertex to the other through a bigger number of petals which,
by definition constitute the {\bf non-parental} petals of the flower.
Below, the pictures of figures~\ref{til_mantilla} indicates how a sector attached to a 
flower can be split into sectors or half-sectors also attached to the flowers. 
In order the reader can understand the pictures we sketchily define the sectors:
in an~$F$-flower, we issue a ray from each red-vertex, following an edge of the 
non-parental petal sharing the red-vertex which does not meet the centre. The lines which
continue the two rays meet in the mid-point of the tile~$B$ in the picture of an~$F$-flower,
in figure~\ref{til_mantilla}. The sector is defined by the angular sector of these two rays.

\vskip 15pt
\setbox110=\hbox{\epsfig{file=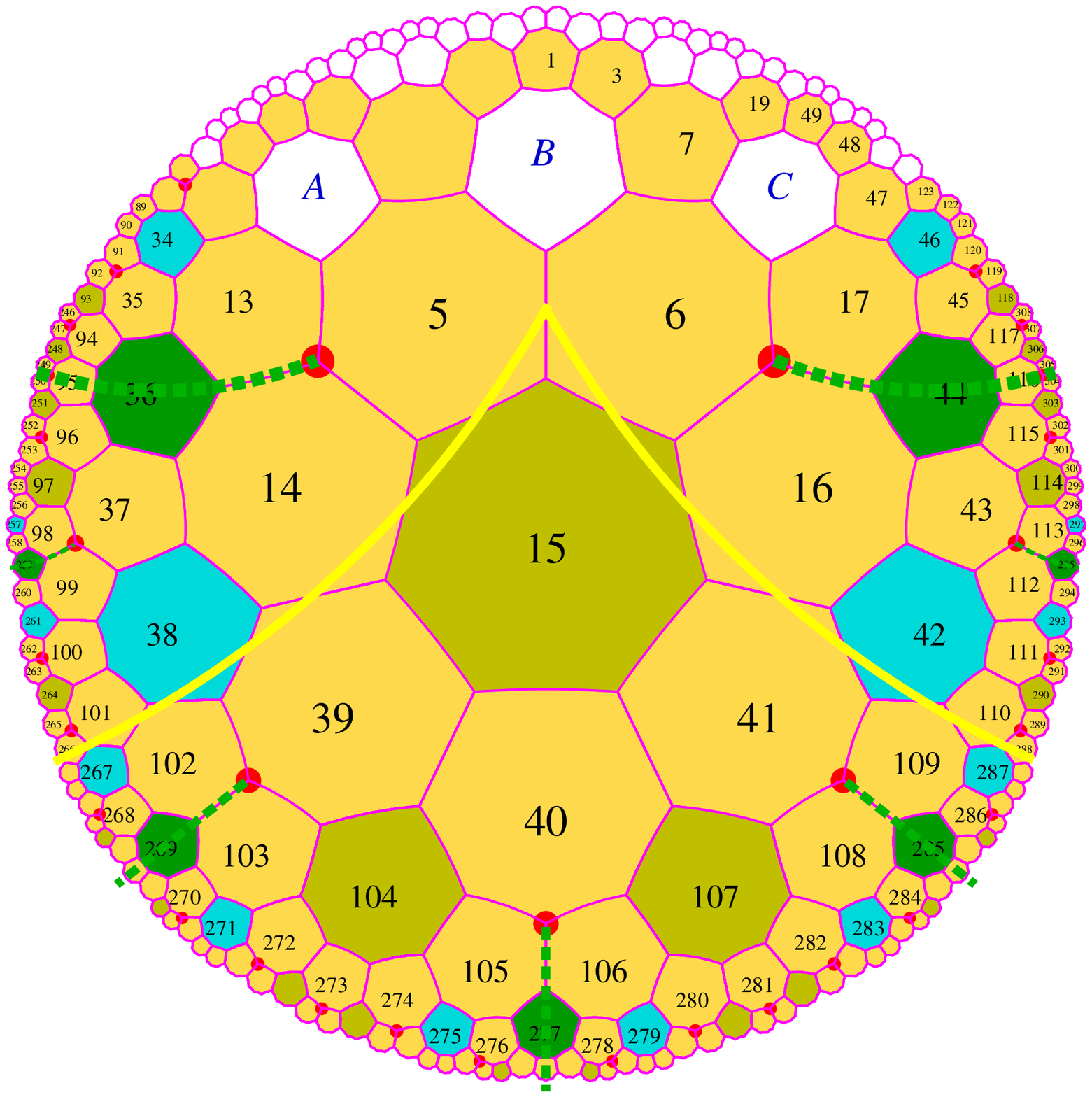,width=100pt}}
\setbox112=\hbox{\epsfig{file=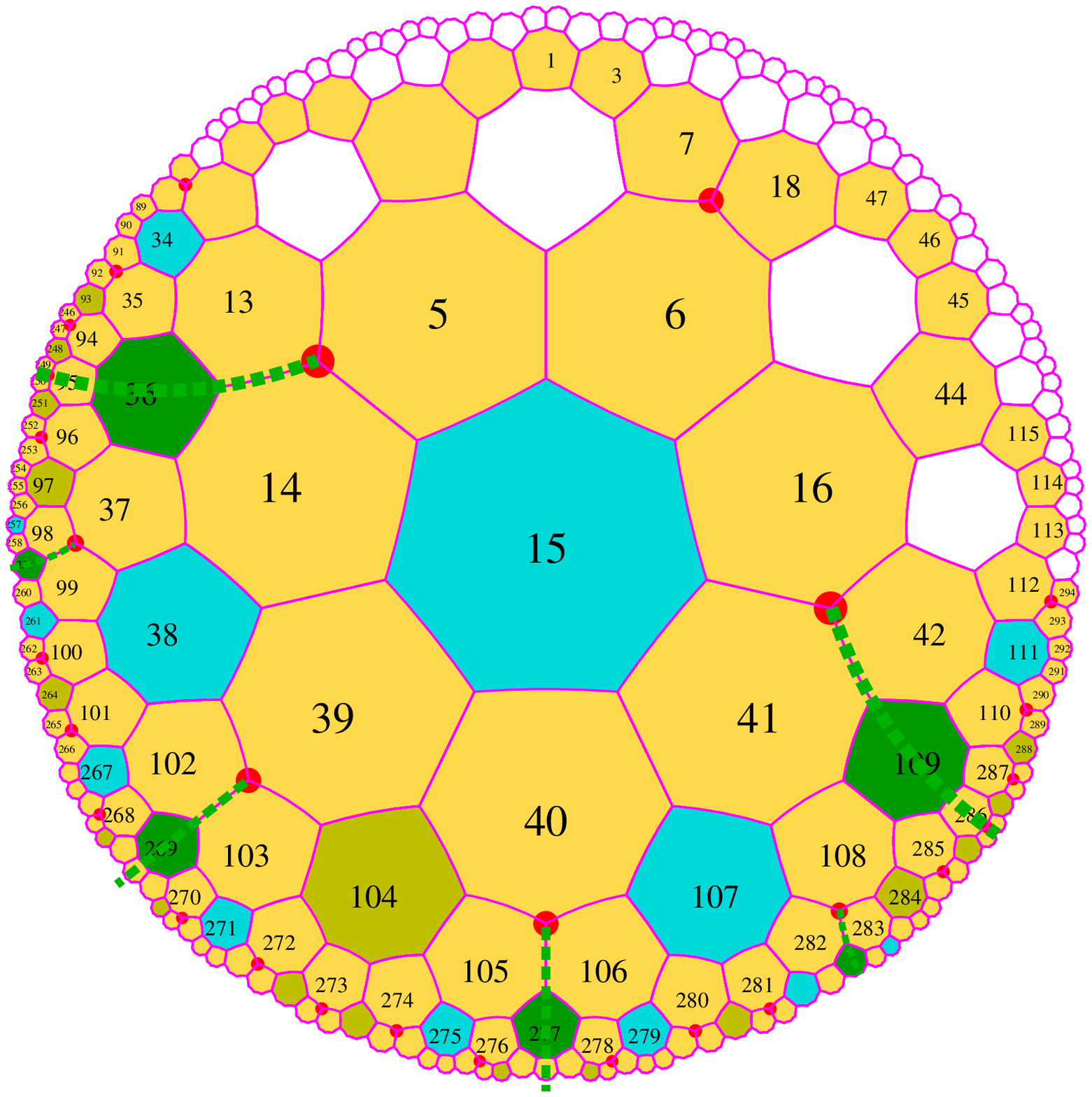,width=100pt}}
\setbox114=\hbox{\epsfig{file=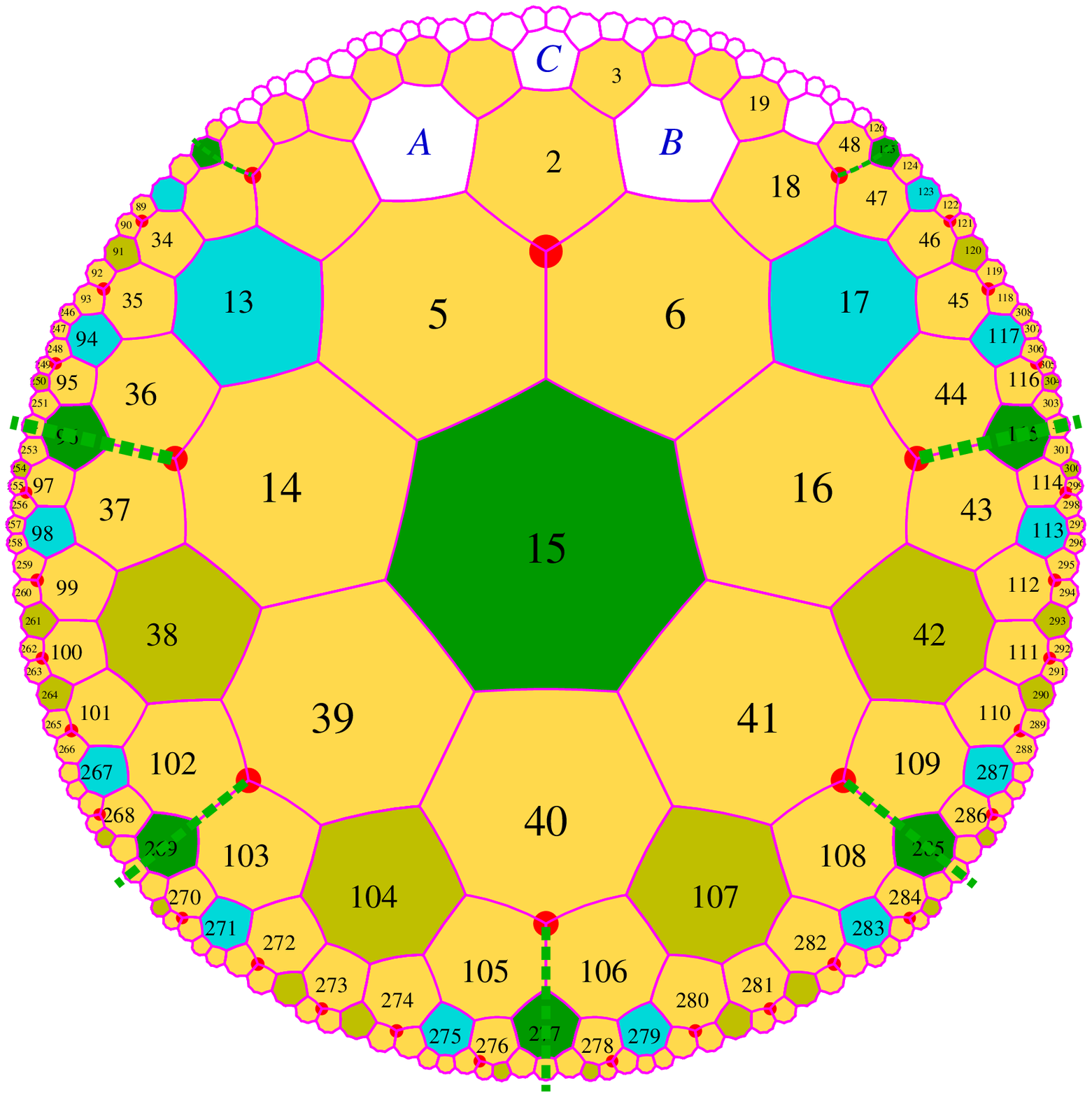,width=100pt}}
\vtop{
\ligne{\hfill
\PlacerEn {-345pt} {-60pt} \box110
\PlacerEn {-255pt} {-55pt} {$F$}
\PlacerEn {-225pt} {-60pt} \box112
\PlacerEn {-135pt} {-55pt} {$G$}
\PlacerEn {-105pt} {-60pt} \box114
\PlacerEn {-15pt} {-55pt} {\bf 8}
}
\begin{fig}\label{til_mantilla}
\leurre
Splitting of the sectors defined by the flowers. From left to right:
an $F$-sector, a $G$-sector and an {\bf 8}-sector.
\end{fig}
}

   In a $G$-flower, we define the sector in a similar way, as the continuations
of the ray meet in the centre of the flower, this time. The sector is the one which is
defined by the rays and which contains an $F$-flower which touches the $G$-centre.
In an {\bf 8}-flower, the sector is the union of the four $F$-sectors defined by
the four $F$-flowers around the centre. These figures induce rules from which we
obtain a recursive process which produces the tiling. In a sector, such a process
defines a tree. The nodes are the centres and the sons of a node are the centres
of the sectors into which the main sector is split.

   Based on these considerations, we have the following result which is
thoroughly proved in \cite{mmtechund,mmTIA}.

\begin{lem}\label{mantilla_tiles} 
There is a set of $4$~tiles of type {\bf centre} and $17$~tiles of type {\bf petal}
which allows to tile the hyperbolic plane as a mantilla. Moreover, there is 
an algorithm to perform such a construction.
\end{lem}

   Next, we define the isoclines, which are obtained in the following way, based on the
following property. Define the status of a tile as~{\bf black} or~{\bf white}, applying 
the usual rules of such nodes in a Fibonacci tree, see \cite{mmbook1}. Then, we have:

\begin{lem}\label{blackseed}
It is possible to require that {\bf 8}-centres are always black tiles. When
this is the case, the $F$-son of a $G$-flower is always a black tile.
\end{lem}

\vskip 5pt
\setbox110=\hbox{\epsfig{file=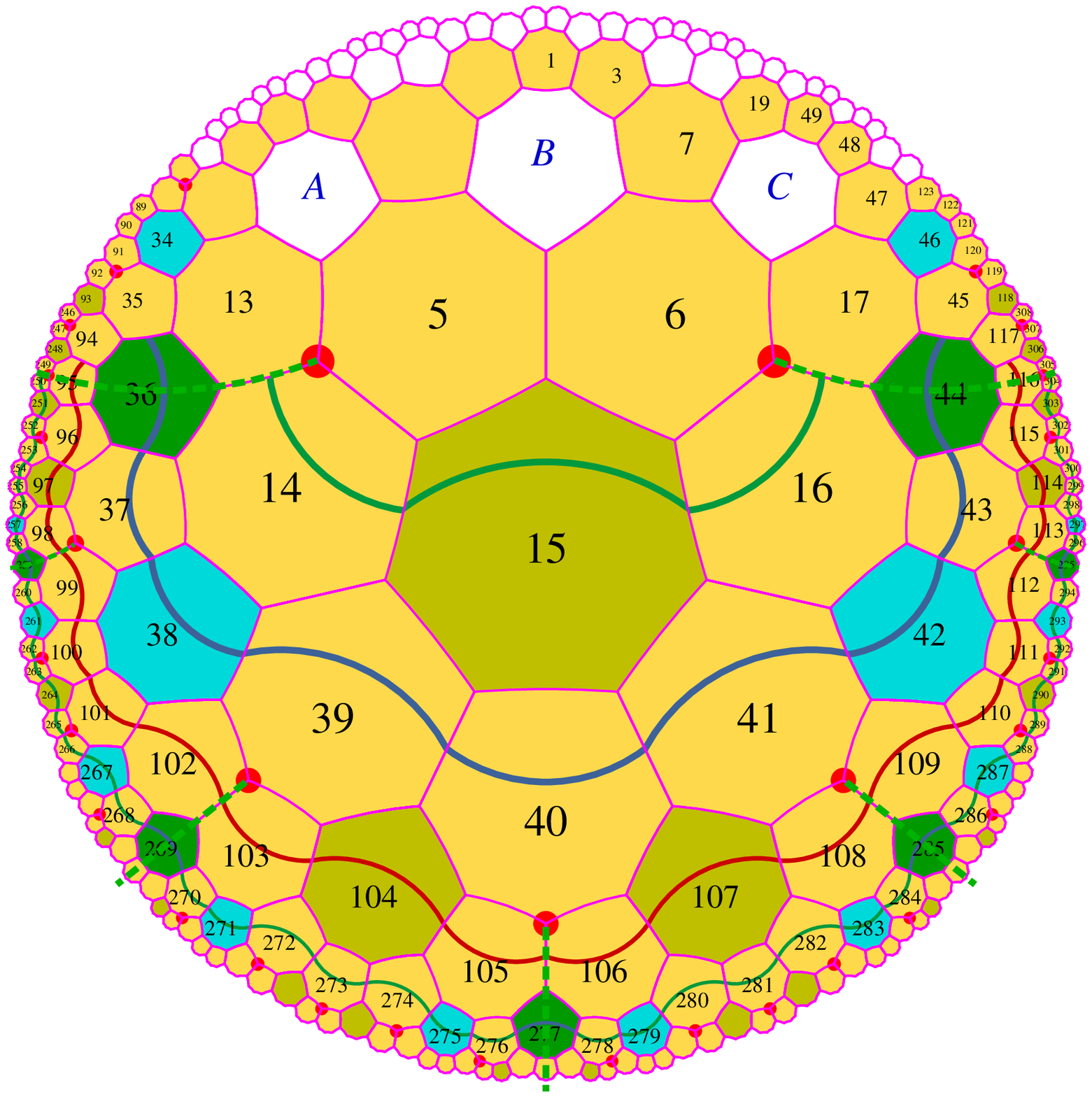,width=110pt}}
\setbox118=\hbox{\epsfig{file=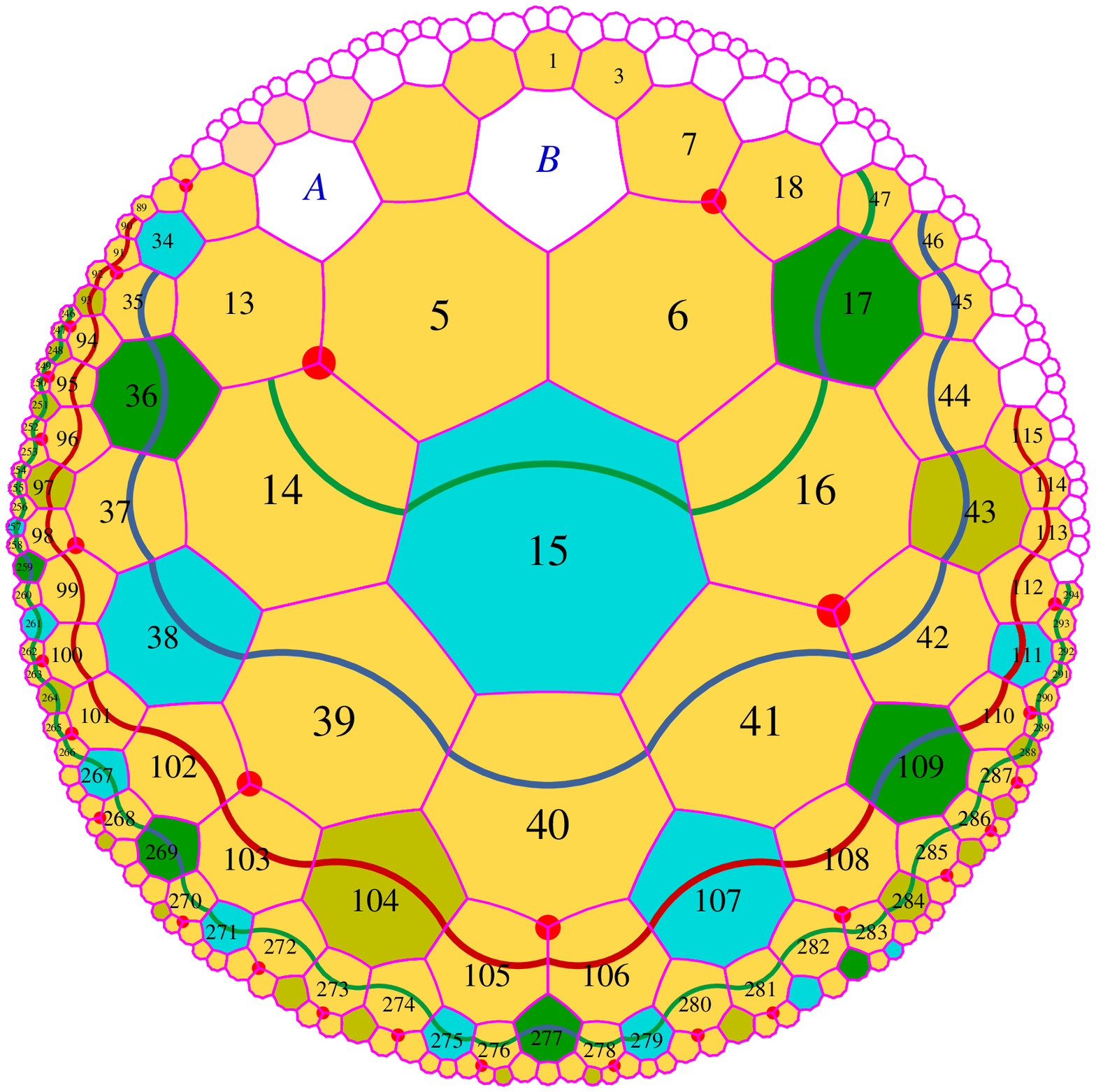,width=110pt}}
\vtop{
\ligne{\hfill
\PlacerEn {-330pt} {0pt} \box110
\PlacerEn {-170pt} {0pt} \box118
}
\begin{fig}\label{blackseedfig}
\leurre
The black tile property and the levels:
\vskip 0pt\noindent
On the left-hand side, a black $F$-centre; on the right-hand side,
a black $G_\ell$-centre. We can see the case of an {\bf 8}-centre on both
figures.
\end{fig}
}

Define arcs as follows, see~\cite{mmnewtechund}: in a white
tile, the arc joins the mid-point of the sides which have a common vertex with 
the side to the father. In a black tile, the arc joins the mid-point of 
the sides separated by the side to the father and the side to the uncle, 
on the left-hand side of the father. Joining arcs, we get paths. The maximal
paths are called {\bf isoclines}. They are illustrated on 
figure~\ref{blackseedfig}. An isocline is infinite and it splits the hyperbolic
plane into two infinite parts. The isoclines from the different trees match, 
even when the areas are disjoint.

   We number the isocline from 0 to 19 and repeat this, periodically.
This defines {\bf upwards} and {\bf downwards} in the hyperbolic plane.
By definition, we call {\bf seed} the $F$-sons of $G$-flowers which are on an
isocline~0, 5, 10 or~15. The reason of this choice lies in the following property:

\begin{lem} \label{iso5} {\rm(see \cite{mmnewtechund,mmBEATCS})} $-$
Let the root of a tree of the mantilla~$T$ be on the isocline~$0$.
Then, there is a seed in the area of~$T$ on the isocline~$5$.
If an {\bf 8}-centre~$A$ is on the isocline~$0$, starting from the
isocline~$4$, there are $F$-sons of $G$-flowers on all the levels. From the
isocline~$10$ there are seeds at a distance at most~$20$ from~$A$.
\end{lem}

   Now, we look at the implementation of the interwoven triangles defined in the
Euclidean plane, in the hyperbolic plane. 

   The implementation meets two kinds of difficulties: first, we have to define how to
implement the triangles and then, the consequences of this definition on the implementation
itself. 

The seeds will be candidate for vertices of trilaterals: any vertex will be defined by a seed
but, not any seed will be allowed to define a vertex. 

   We decide that {\bf all} the seeds of an isocline~0 define the vertex of a blue-0
{\bf triangle}, we say that the seeds of an isocline~0 are {\bf active}. Now, an active seed
defines a Fibonacci tree whose legs are supported by the borders of the tree: its left- and
its rightmost branches, see \cite{mmBEATCS,mmarXiva}. 
The seed dispatches a {\bf scent} which propagates to each tile, inside the tree and on 
its borders too, down to the fifth level after the level of the seed. If the tile which
is met there by the scent is not a seed, the scent dies at this tile. If the tile
met by the scent there is a seed, then this seed becomes {\bf active} to its turn.
Now, the bases of the triangles of the generation~0 are defined by the active
seeds of the isoclines~10 which, at the same time, issue the legs of the phantoms of
the generation~0. The scent issued from the active seeds of the isoclines~10 also
raises the green line which runs on the mid-point line of the phantoms of the generation~0
which is supported by an isocline~15. This fully defines the trilaterals of the
generation~0. 

   Next, the scent which meets an isocline~15 also emits the green line which is now called
the {\bf green signal}. It is used for the construction of the further generations.
The vertices of the trilaterals of the generation~1 occurs on the isoclines~5. And next,
for all the other generations, the vertex is defined by a seed which is activated on
an isocline~15. The process is mainly the same as in the Euclidean case.

   However, two concurrent phenomena occur which might disturb the construction: 
the same isocline may cross several triangles inside a given triangle; the same isocline 
may cross trilaterals of the same generation as well as trilaterals of a smaller generation
inside a trilateral of a bigger one or in between two contiguous triangles of a 
bigger generation. As an example, on the same isocline, we may have a green signal inside
two contiguous triangles of the same generation as well as, in between, the green signal
of smaller phantoms. Now, the phantoms do not stop the green signal while the triangles do.

   Here, we do not have the room to indicate how to handle this situation. This is 
performed in \cite{mmBEATCS,mmarXive} under the name of {\bf synchronization mechanisms} 
which are introduced to guarantee the correctness of the construction algorithm which was
devised for the Euclidean implementation. What is important here is that {\bf we have 
an implementation of the interwoven triangles in the hyperbolic plane}.
  
   An another important property of this implementation is the density of the active seeds
in the hyperbolic plane:

\begin{lem}\label{density} {\rm (see \cite{mmBEATCS,mmarXiva,mmnewtechund})} $-$
For any tile~$\tau$ of the mantilla, there is an active seed within a ball of 
radius~$20$ around~$\tau$.
\end{lem}
 
\subsection{An application to the tiling problem}

   In fact, this construction was used to prove the following result:

\begin{thm}\label{GTP} {\rm (Margenstern, \cite{mmBEATCS,mmarXive}, Kari, \cite{jkariMCU})} 
$-$
The tiling problem is undecidable for the hyperbolic plane.
\end{thm}
 
   The two indicated proofs are completely different and independent. Also, 
\cite{mmarXiva,mmarXive} occurred first. Also, the proof given in 
\cite{mmBEATCS,mmarXiva,mmarXive} has the following property: given the 1-bit information
that the considered finite set constructed for the proof tiles the plane, the proof
provides an algorithm which constructs the tiling, of course, in infinite time. 
It is interesting to note that the interwoven triangles define a strongly non-periodic
tiling of the hyperbolic plane, as Chaim Goodman-Strauss pointed to me. 
Remember that, in the case of the Euclidean plane, Berger's proof of the undecidability
of the same tiling problem for this plane, see~\cite{berger},  was the first example of 
an aperiodic tiling. However, in the case of the hyperbolic plane, the existence of 
strongly aperiodic set of tiles was known before the solution to the result stated by 
theorem~\ref{GTP}, see~\cite{goodman}.
 
   As we need this construction of the interwoven triangles for the result of this paper, 
we sketchily mention how we derive the proof of theorem~\ref{GTP} from the previous 
two sub-sections.

   First, we ignore the blue-0 and the blue triangles, the phantoms of 
any colour as well as most of the construction signals indicated in \cite{mmBEATCS}
for instance. We just keep the {\bf yellow signal} which, by definition, marks the free
rows in the red triangles. Indeed, the free rows of the red triangles constitute the 
horizontals of the grid which we shall construct in order to simulate the space-time 
diagram of a Turing machine. 

   Now, we have to define the verticals of the grid to complete the 
simulation. They consist of rays which cross {\bf 8}-centres. 
Figure~\ref{vertical_1} illustrates the various ways of their connection with
the tile of a border of the tree being on a free row.

   The computing signal starts from a the seed. It travels on the free rows. 
Each time a vertical is met, which contains a symbol of the tape, the required 
instruction is performed. If the direction is not changed and the corresponding
border is not met, the signal goes on, on the same row. Otherwise, it goes down
along the vertical until it meets the next free row. There, it looks at the 
expected vertical, going in the appropriate direction. Further details are 
dealt with in \cite{mmnewtechund}.

\vskip 15pt
\setbox110=\hbox{\epsfig{file=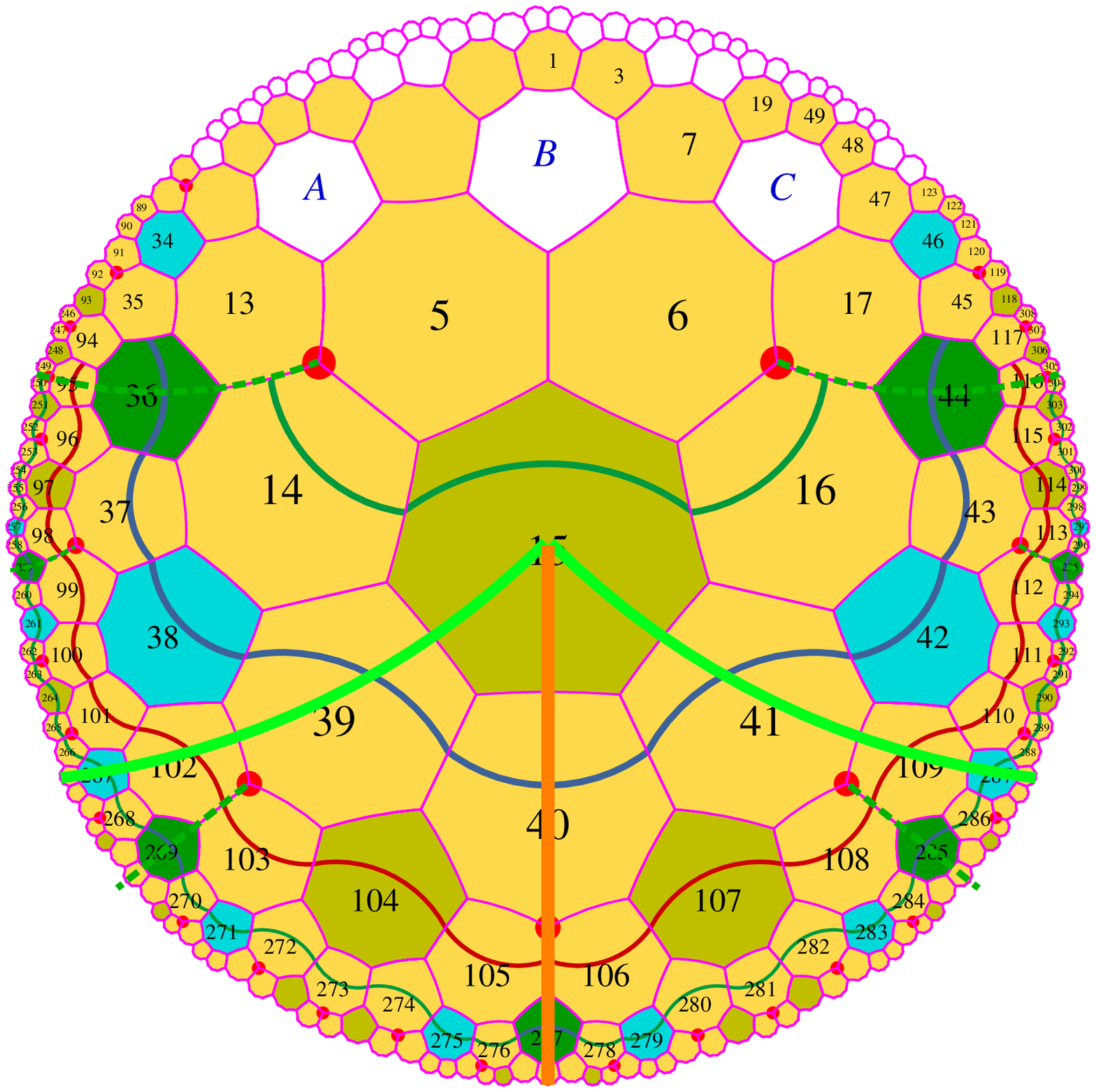,width=110pt}}
\setbox112=\hbox{\epsfig{file=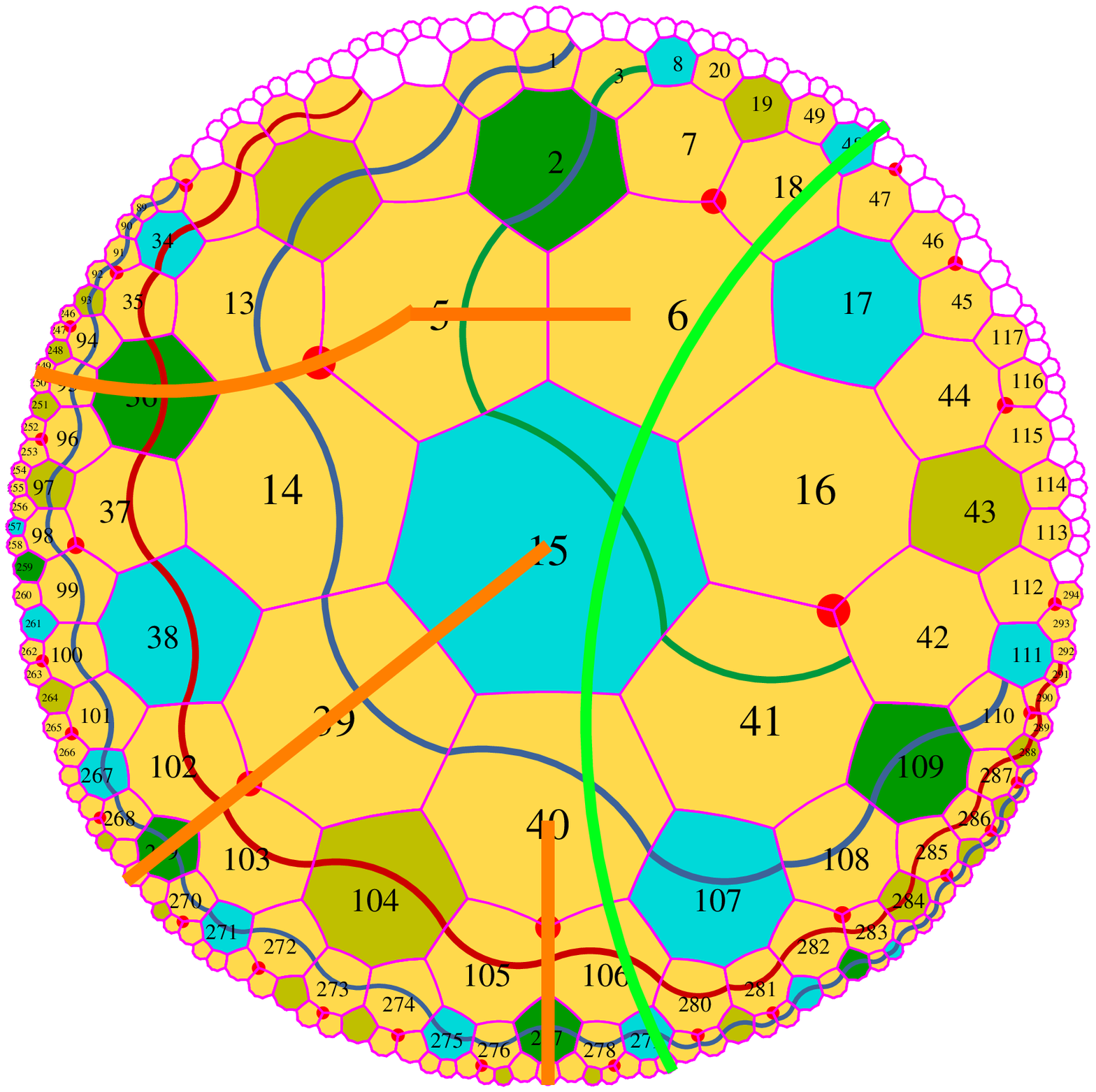,width=110pt}}
\ligne{\hfill
\PlacerEn {-130pt} {0pt} \box110
\PlacerEn {20pt} {0pt} \box112
\hfill}
\vskip-15pt
\begin{fig}\label{vertical_1}
\leurre
The perpendicular starting from a point of the border of a triangle which
represents a square of the Turing tape.
\vskip 0pt
On the left-hand side: the case of the vertex. On the right-hand side,
the three other cases for the right-hand side border are displayed on the
same figure.
\end{fig}

   As indicated in \cite{mmBEATCS,mmarXiva,mmarXive}, there are continuously many
realizations of the mantilla and also continuously many realizations of the interwoven
triangles in each realization of the mantilla. One realization of the interwoven
triangles will keep our attention: it is what we have called the {\bf butterfly model},
in \cite{mmBEATCS,mmarXiva,mmnewtechund}. In this model, there is a single isocline~15
which is never cut by red triangles. This isocline may be the basis of an infinite red 
phantom and, in this case, it contains infinitely many vertices of red triangles which, 
therefore, are infinite.

\section{The mauve triangles and a plane filling path}

   In \cite{mmCSJMfill}, we defined a construction which, up to a point, forces the
construction of a plane filling path in the hyperbolic plane.

   Here, let us remember the main lines of this construction.

\subsection{The mauve triangles}

   The first step of the construction is the introduction of the {\bf mauve triangles}.
We again start from an implementation of the interwoven triangles in the hyperbolic
plane. The reason of these new triangles will become more clear after the next
sub-section where we construct a path which crosses each tile of the tiling exactly once. 
At the present moment, we can simply say that these new triangles allow to better control
the various turns of the path in order to satisfy the just indicated constraint.  
   
   As indicated in~\cite{mmCSJMfill}, the mauve triangles are constructed from the 
red triangles. They double the height of the red triangles: accordingly, they
loose the separation property of the red triangles. Mauve triangles do intersect inside
a generation and also in between generations: this point raises several difficulties which
are overcome as will be indicated below.

   Each vertex of a red triangle is also the vertex of a mauve triangle, and conversely.
Let~$V$ be a vertex of a red triangle~$R$. The legs of the mauve triangle~$M$ with
vertex~$V$ are supported by the same branches of the Fibonacci tree rooted at~$V$ 
as~$R$. Now, the legs of~$M$ go further on the branches. They continue until they
meet the next isocline which contains the vertices of red triangles of the same generation
as~$R$. And so, the height of~$M$ is twice the height of~$R$. The construction of the mauve
triangles from the red triangles supporting them is not very difficult. The reader is
referred to~\cite{mmCSJMfill,mmarXivf} for useful details and constructions.

   The mauve triangles have interesting properties which are stated by the following two
lemmas, taken from~\cite{mmCSJMfill}. We call {\bf latitude} of a mauve triangle the set
of isoclines which crosses the legs of the triangle from its vertex to its basis, the
isoclines of the vertex and of the basis being included.

\begin{lem}\label{cover}
Let~$\tau$ be a tile of the tiling. Then for any non-negative~$n$,
there is a mauve latitude $\Lambda$ of the generation~$n$ such
that $\tau\in\Lambda$. And then: either $\tau$ falls within a mauve
triangle of generation~$n$ in this latitude or $\tau$ falls outside
two consecutive mauve triangles of generation~$n$ and of the
latitude~$\Lambda$ and in between them.
\end{lem}

\vskip-40pt
\setbox110=\hbox{\epsfig{file=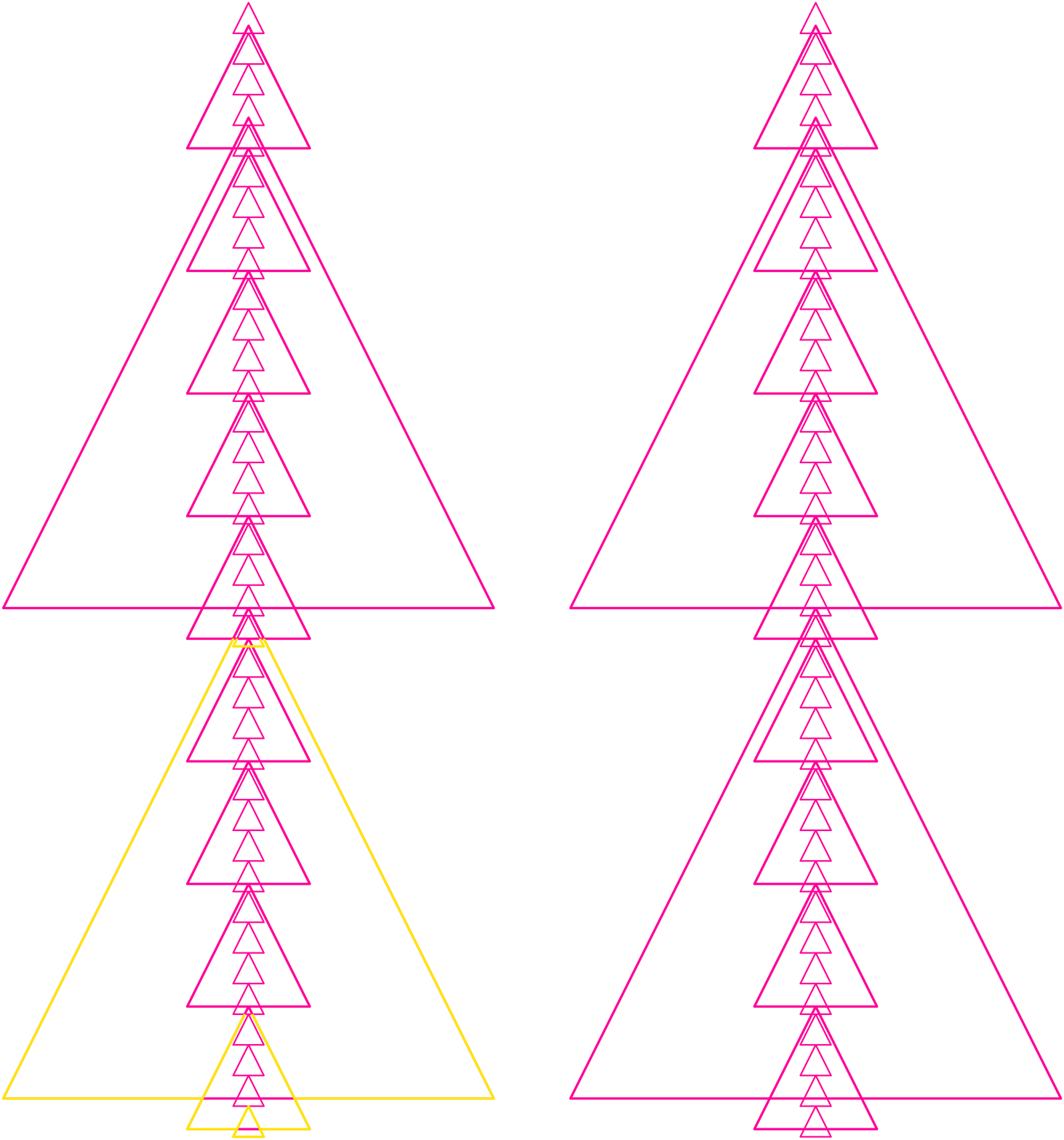,width=250pt}}
\ligne{\hfill
\PlacerEn {-130pt} {0pt} \box110
\hfill}
\vspace{-5pt}
\begin{fig}\label{les_mauves}
\leurre
An illustration of the mauve triangles. 
\end{fig}
 
   This property follows immediately from the fact that the latitude of
a mauve triangle exactly covers that of the corresponding red triangle
and the following latitude of red phantoms.

   However, there is a price to pay to this. Remember: the red triangles are
either disjoint or embedded. Now, mauve triangles do intersect from one
generation to another. Fortunately, this intersection is not that big
and it is characterized by the following statement.

\begin{lem}\label{overlap}
A mauve triangle~$T$ of positive generation~$n$ intersects mauve triangles
of generation~$n$$-$$1$, and it possibly intersects one mauve triangle
of generation~$n$$+$$h$$+$$1$, with $h\geq 0$. When the intersection occurs, 
the legs of~$T$ cut the basis of the mauve triangle of the higher
generation at a point which is on the mid-distance line of the red phantoms 
which share their basis with that of~$T$. Call {\bf low point} this point on the legs
of~$T$. The basis of~$T$ is cut by the legs of mauve triangles of
generation~$m$, with $m<n$, at their low points.
\end{lem}

   In \cite{mmCSJMfill}, we indicate how to construct the low points of the mauve
triangles. We mean by this that the tiles must force the detection of the low points
at the same time as they force the construction of the mauve triangles. This is not 
very difficult to obtain.
 
\subsection{A plane filling path}

   Once the mauve triangles are installed, it is possible to define a path which visits
all the tiles exactly once for each of them. Such a path will here be called a 
{\bf plane filling path}.

   The construction consists in defining a way to fill the mauve triangles as well as
the space in between two consecutive mauve triangles of the same generation and within the
same latitude. This latter space is called a {\bf trapeze}.

   First, we split the mauve triangles and the trapezes in slices which again involve
mauve triangles and trapezes of a smaller generation. Then, we shall proceed to
a tuning: unfortunately, things cannot be that simple, due to the property stated
by lemma~\ref{overlap} about the possible overlapping of mauve triangles of different 
generations. Such a splitting is suggested by figure~\ref{tri_trap}.

   Figure~\ref{quadrangle_deux} indicates when the slicing is refined to a second smaller
generation.

\setbox110=\hbox{\epsfig{file=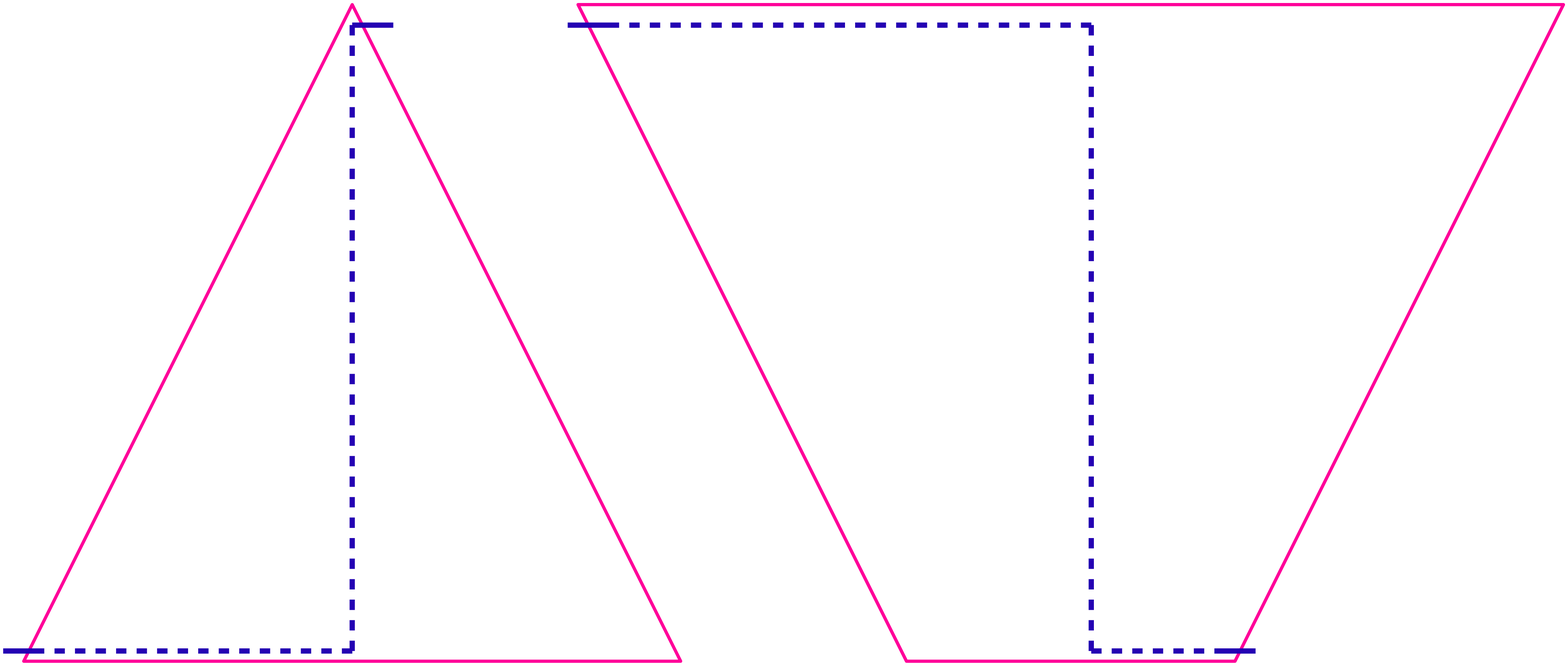,width=180pt}}
\setbox112=\hbox{\epsfig{file=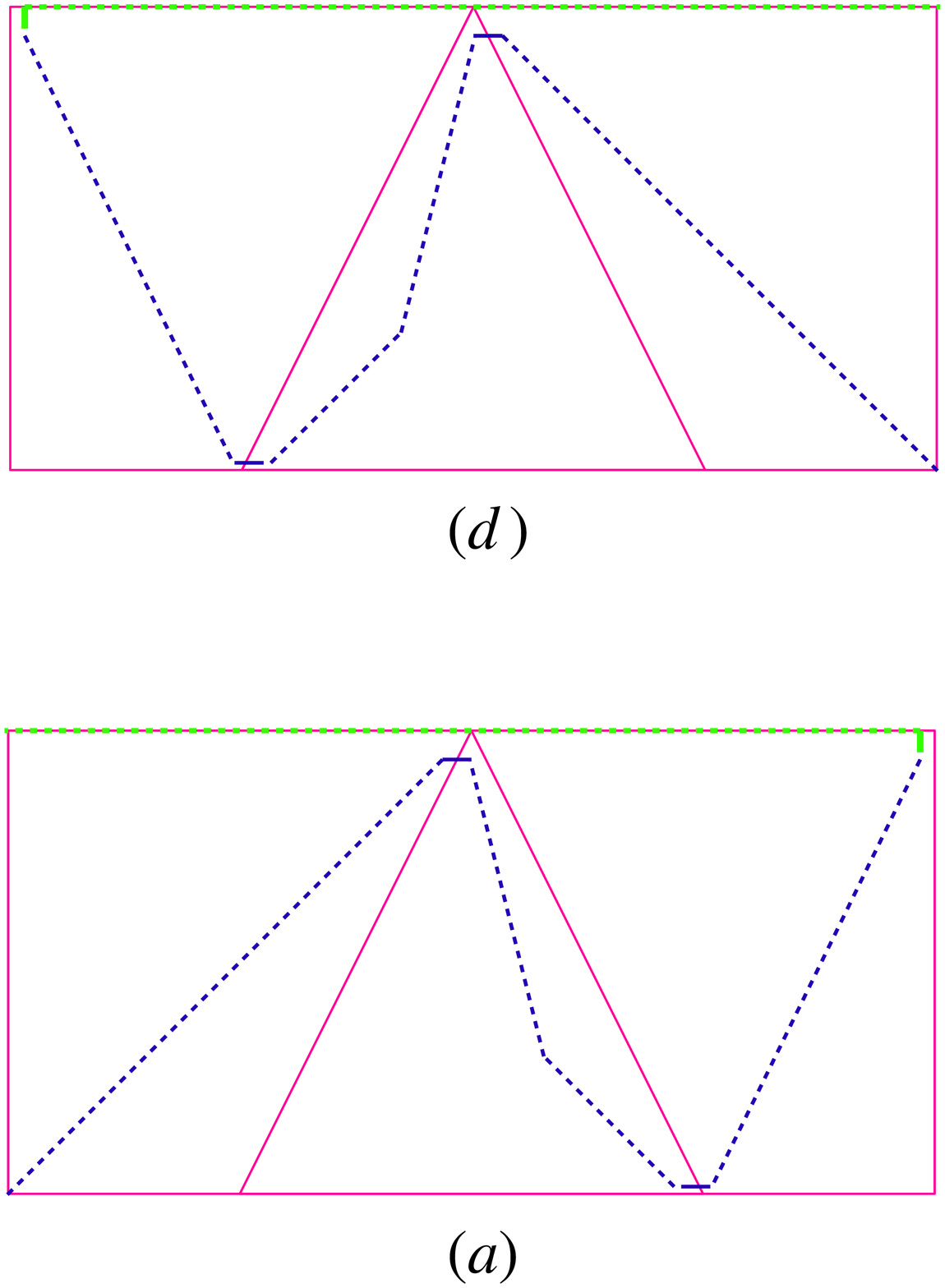,width=90pt}}
\vskip 10pt
\vtop{
\ligne{\hfill
\PlacerEn {-330pt} {-10pt} \box110
\PlacerEn {-130pt} {-40pt} \box112
}

\begin{fig}\label{tri_trap}
\leurre
The basic figures: triangle and trapeze within a latitude, and the splitting of the slices.
\vskip 0pt\parindent 0pt
On the right-hand side: above, the descending case; below, the ascending case.
\end{fig}
}

\setbox110=\hbox{\epsfig{file=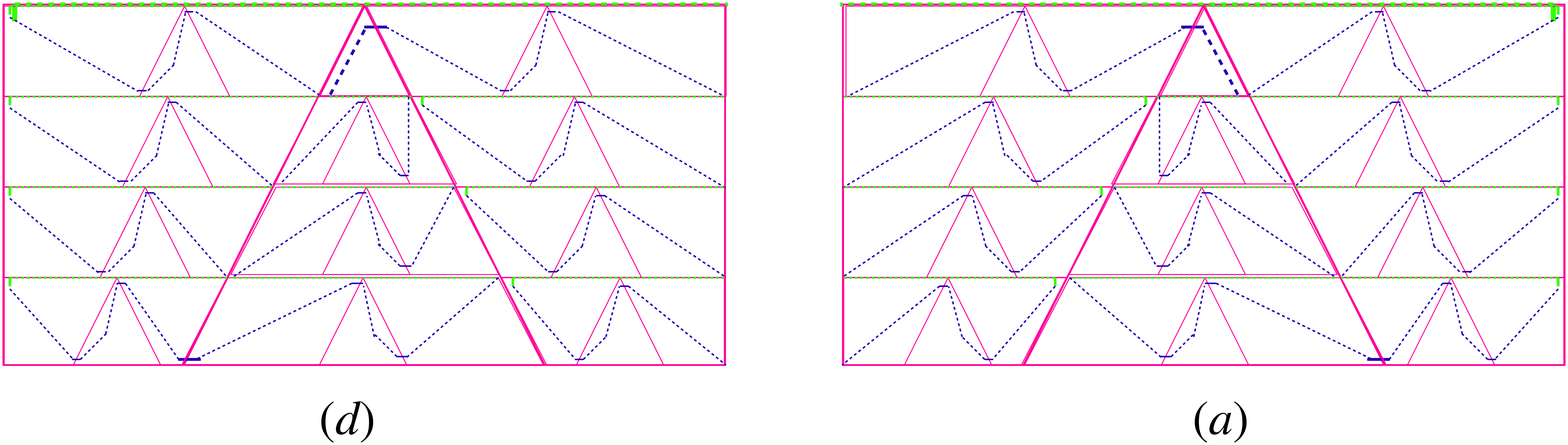,width=350pt}}
\vskip -15pt
\vtop{
\ligne{\hfill
\PlacerEn {-350pt} {-40pt} \box110
}

\vspace{-30pt}
\begin{fig}\label{quadrangle_deux}
\leurre
The splitting of the slices: second generation.
\vskip 0pt\parindent 0pt
On the left-hand side: the descending case. On the right-hand side:
the ascending case.
\end{fig}
}

\vskip 7pt
   At this point, it should be indicated that the situation is not as simple as it can
be viewed from the figures~\ref{tri_trap} and~\ref{quadrangle_deux}. Indeed, in these
figures, the representation ignores the fact that mauve triangles overlap. As can be seen
in figure~\ref{les_mauves}, if we take into account the overlap, we have to change the
definition of the basis of a mauve triangle, as well as its vertex.

   As indicated in \cite{mmCSJMfill,mmarXivf}, we re-define the bottom a mauve triangle.

First, the entry point into a triangle is not necessary one of its corner. It may be
the low point. This depends on the fact whether the mauve triangle is crossed or not
by a triangle of a bigger generation. As can be deduced from figure~\ref{les_mauves}
this may happen and it may also not happen as well. When it happens, the basis of the
bigger triangle cuts the legs of the smaller one at their low points.

   In the case when the mauve triangle~$T$ is not crossed by a bigger one, the entry is
at a corner and the basis follows the lower part of the inner triangles whose legs are
cut by the basis of~$T$. Now, for these lower parts, we apply recursively the same principle.
When the mauve triangle~$T$ is crossed by a bigger one, the entry is at the low point: other
wise, there would be a contradiction between the path following the basis of the
big triangle and that which runs inside~$T$. As explained in \cite{mmCSJMfill}, 
the path follows the lower part of~$T$,  which is also in agreement with what we have just
defined above, and we apply the same recursive process for further intersections down to
a mauve triangle of the first generation.

   We call the new definition of this bottom of a mauve triangle, the {\bf refined
basis}. This correspondingly may redefine the vertex of a mauve triangle~$T$, as the 
upper part of the legs of~$T$, as it may be cut by a refined basis. 
See~\cite{mmCSJMfill,mmarXivf} for more precise information. With these modifications, 
the above indications define a path which passes through all tiles exactly once. Moreover, 
this path has no origin in this sense that no tile play a special r\^ole with respect to 
the path. We shall say that the path is a {\bf uniform plane-filling path}.

   Accordingly, we have proved the following theorem:

\begin{thm}\label{uniplanefill} {\rm (Margenstern, see \cite{mmCSJMfill})} $-$
There is a uniform plane-filling path for the ternary heptagrid 
in hyperbolic plane.
\end{thm}

\subsection{The exceptional case: the butterfly model}

   However, this construction does not provide us with a {\it strong} plane filling
property as the construction of Jarkko Kari does for the Euclidean plane.

   In fact, when the realization of the interwoven triangles has the property that
any tile falls in the interior of some mauve triangle, then our construction provides
us with a uniform plane filling path which is then forced by the tiling.

   But there is a realization when this may be not the case: it is the situation of the
butterfly model.
\vskip 7pt
   When the realization of the interwoven triangles is the butterfly model, there
can still be four cases, depending on the basis which accompanies the infinite green line.
There is one case where our construction does not provide a plane filling path: it is the
situation when the infinite green line is accompanied by the basis of a red phantom.
In this case, this basis can be the basis of an infinite mauve triangle, of course,
applying to it the refinements which we have mentioned in the previous sub-section.

\section{About the injectivity of the global function of a cellular automaton in the
hyperbolic plane}

   If we had not this exceptional situation, we could mainly apply the argument 
of~\cite{jkari94}, with a slight modification. Remember that the automaton~$A_T$
attached to a set of tiles~$T$ in~\cite{jkari94} has its states 
in $D\times\{0,1\}\times T$, where $D$ is the set of tiles which defines the tiling 
with the plane filling property and~$T$ is an arbitrary finite set of tiles. 
We can still tile the plane as we assume that the tiles of~$T$ are ternary heptagons
but the abutting conditions may be not observed: if it is observed with all the neighbours
of the cell~$c$, the corresponding configuration is said to be {\bf correct} at~$c$,
otherwise it is said {\bf incorrect}. When the considered condiguration is correct at every
tile for~$D$ or at every tile for~$T$, it is called a {\bf realization} of the corresponding
tiling. Let~$\delta$ denote the path induced by a realization of~$D$.  As in~\cite{jkari94}, 
the transition function does not change neither the $D$- nor the \hbox{$T$-component} of the 
state of a cell~$c$: it only changes its $\{0,1\}$-component~$x(c)$ which is later on 
called the {\bf bit} of~$c$. As in~\cite{jkari94}, we define $A_T(x(c))=x(c)$ if the 
configuration in~$D$ or in~$T$ is incorrect at the considered tile. If both are correct, 
we define $A_T(x(c))=\hbox{\rm xor}(x(c),x(\delta(c)))$. It is plain that if~$T$ tiles the 
hyperbolic plane, there is a configuration of~$D$ and one of~$T$ which are realizations
of the respective tilings. Then, the transition function computes the xor of the bit
of a cell and its next neighbour along the path. Hence, defining all cells with~0 
and then all cells with~1 define two configurations which~$A_T$ transform to the same
image: the configuration where all cells have the bit~0. Accordingly, $A_T$ is not injective.

Conversely, if $A_T$~is not injective, we have two different configurations~$c_0$ and~$c_1$
for which the image is the same. Hence, there is a cell~$t$ at which the configurations
differ. Hence, the xor was applied, which means that~$D$ and~$T$ are both correct at this
cell in these configurations and it is not difficult to see that the value
for each configuration on the next neighbour along the path must also be different.
And so, following the path in one direction, we have a correct tiling for both~$D$ and~$T$.
Note that this entails nothing for the part of the path which is before~$t$.
But we need not to take {\bf any} finite set of tiles. We can take the family of
set of tiles defined in~\cite{mmBEATCS} which are attached to Turing machines, as
sketchily mentioned in subsection~2.3. We denote by~$T_M$ such a finite set of tiles 
attached to a Turing machine~$M$. Indeed, it is not difficult to see that the half of 
the path after~$t$ covers infinitely many mauve triangles of arbitrary sizes, hence it 
covers infinitely many balls of arbitrary sizes. Now, in these balls, $T_M$ is also 
correct and so, by the density property of the seeds, there are infinitely many red 
triangles of~$T_M$ which are correct and so, the computation of~$M$ does not halt. And 
so, we can see that the injectivity of~$A_{T_M}$ would be reduced to the halting problem.

   Now, as indicated, this cannot be derived as there is an exceptional situation
in which we could have a non injective cellular automaton although the Turing machine
halts: it is enough to take the above mentioned configuration with the infinite mauve
triangle in which $T_M$ is correct along the refined basis of the infinite mauve triangle.

   But, we can impose a condition which boils down to ignore the refined basis.

   Say that $A$ is a {\bf $\beta$-freezed cellular automaton}, if $A$ is associated 
with a fixed refined basis~$\beta$ of a mauve triangle and if its transition function if
the identity on all cells which belong to~$\beta$. We also say that $A$~is {\bf freezed
along} $\beta$. Then we have:

\begin{thm}\label{undecinjfreeze}
The injectivity problem is undecidable for $\beta$-freezed cellular automata on
the ternary heptagrid.
\end{thm}

\noindent
Proof. Indeed, we define $D$~and~$T_M$ as previously. The set of tiles~$D$ defines
a uniform plane filling path in the conditions of theorem~\ref{uniplanefill} and
$T_M$ is associated to the Turing machine~$M$ as in~\cite{mmBEATCS}. We define
the transition function of a $\beta$-freezed cellular automaton as described above, 
except when the cell is on~$\beta$: in that case, necessarily, the new state is the same
as the previous one.

Assume that~$M$ does not halt. Then, there is a configuration for~$D$ which corresponds
to the realization of~$D$ as a butterfly model with~$\beta$ as the refined basis of
an infinite mauve triangle. There is also a configuration for~$T_M$ which is a realization
of a tiling of the hyperbolic plane by~$T_M$. Then $A_{T_M}$ is not injective: take
again the configuration where all cells have the bit~0 and the configuration when they
all have the bit~1 except on~$\beta$ where the cells have the bit~0. As the xor function 
is applied, and as the path is always defined, even when it is not unique, we get 
that $A_{T_M}$ is not injective: both indicated configurations are transformed in the
configuration where all the cells have the bit~0.

   Now, assume that $A_{T_M}$ is not injective. In this case, there are two different
configurations $c_0$ and $c_1$ which have the same image. Necessarily there is a tile~$t$
such that $c_0(t)\not=c_1(t)$. As the automaton is freezed along~$\beta$, $t\not\in\beta$.
And so, either~$t$ is above~$\beta$, either it is below. In bot cases, we can apply the
argument which we have above reproduced from~\cite{jkari94}, with this restriction
that only the path after~$t$ is correct as well as the tiling of~$T_M$ along this 
restriction to a half of the path.

   Now, as indicated in~\cite{mmCSJMfill}, in the exceptional situation of the butterfly
model with a refined basis of an infinite mauve triangle, above~$\beta$ there is a
single path defined by~$D$ which visits all the tiles above~$t$ exactly once. And so,
if $t$~is above~$\beta$, we conclude that~$M$ does not halt. Assume that~$t$ is 
below~$\beta$. Then, from~\cite{mmCSJMfill,mmarXivf}, we know that the part of the tiling
which is below~$\beta$ can be split into infinitely many regions in which the path visits
all the tiles exactly once and in which there are infinitely many mauve triangles
with arbitrary sizes. Indeed, we can sketchily describe these regions. In the considered
situation, there are infinitely many infinite mauve triangles whose vertices are on~$\beta$.
Each region is defined by such an infinite mauve triangle~$\cal T$ rooted on~$\beta$ and
the intermediate region between~$\cal T$ and the next infinite mauve triangle~$\cal M$ 
rooted on~$\beta$, $\cal M$ being not included. Now, if the path after~$t$ visits~$\cal T$, 
we are done. If, on the contrary, it visits the intermediate region in between~$\cal T$ 
and~$\cal M$, we are also done: indeed, it is not difficult to see that in this intermediate 
zone, there are infinitely many mauve triangles with arbitrary sizes.

And so, in all cases, the half of the path after~$t$ visits infinitely many mauve 
triangles with arbitrary sizes. And so, the argument which we have above produced allows 
us to conclude that $M$~does not halt.

   Accordingly $M$ does not halt if and only if $A_{T_M}$ is not injective. Accordingly,
the injectivity of the $A_{T_M}$'s is undecidable. \cqfd

\section{Conclusion}

   The question of the injectivity of the global function of cellular automata in the
hyperbolic plane is still open. However, I think that it is undecidable, although the
above argument does not allow to derive this conclusion. Even if the injectivity
happens to be clearly proved as undecidable, we shall remain with the question of the
surjectivity. In the Euclidean case, it is known that the surjectivity of the
global function of a cellular automaton is equivalent to its injectivity on the set
of finite configurations, see~\cite{moore,myhill}. Now, in the case of cellular automata 
in the hyperbolic plane, this is not at all the case. The surjectivity and the 
injectivity of the global function are independent: there are examples of surjective 
global functions which are not injective and of injective global functions which are 
not surjective. Accordingly, this question is completely open in the hyperbolic plane.

%

\end{document}